\documentclass[preprint,12pt]{elsarticle}
\usepackage[left=2.5cm, right=2.5cm]{geometry}
\usepackage{setspace}
\usepackage{ifthen}
\usepackage{amsmath}
\usepackage{amsfonts}
\usepackage{amssymb}
\usepackage{natbib} 

\newtheorem{defin}{Definition}[section]
\newcommand{\definition}[2]{
\begin{defin}#2\label{#1}\end{defin}
}

\newtheorem{rem}{Remark}[section]
\newcommand{\remark}[2][]{
\ifthenelse{\equal{#1}{}} {\begin{rem}#2\end{rem}}
{\begin{rem}#2\label{#1}\end{rem}} }

\newcommand{\bpr}{\textbf{Proof }} 
\newcommand{\epr}{\qquad $ \Box$}    
\newenvironment{proof}[1][]{\textbf{\bpr#1 }\\}{\epr}

\newtheorem{theor}{Theorem}[section]
\newcommand{\theorem}[4][]{
\ifthenelse{\equal{#1}{}} {\begin{theor}#3\label{#2}\end{theor}
\begin{proof}#4\end{proof}}
{\begin{theor}[#1]#3\label{#2}\end{theor}
\begin{proof}#4\end{proof}}
}
\newcommand{\nprtheorem}[3][]{
\ifthenelse{\equal{#1}{}} {\begin{theor}#3\label{#2}\end{theor}}
{\begin{theor}[#1]#3\label{#2}\end{theor}} }
\newcommand{\notheorpr}[2][]{
\ifthenelse{\equal{#1}{}} {\begin{proof}#2\end{proof}}
{\begin{proof}[\mbox{of Theorem }\ref{#1}]#2\end{proof}} }

\newtheorem{prop}{Proposition}[section]
\newcommand{\proposition}[4][]{
\ifthenelse{\equal{#1}{}} {\begin{prop}#3\label{#2}\end{prop}
\begin{proof}#4\end{proof}}
{\begin{prop}[#1]#3\label{#2}\end{prop}
\begin{proof}#4\end{proof}}
}
\newcommand{\nprproposition}[3][]{
\ifthenelse{\equal{#1}{}} {\begin{prop}#3\label{#2}\end{prop}}
{\begin{prop}[#1]#3\label{#2}\end{prop}} }
\newcommand{\noproppr}[2][]{
\ifthenelse{\equal{#1}{}} {\begin{proof}#2\end{proof}}
{\begin{proof}[\mbox{of Proposition }\ref{#1}]#2\end{proof}} }

\newtheorem{lem}{Lemma}[section]
\newcommand{\lemma}[4][]{
\ifthenelse{\equal{#1}{}} {\begin{lem}#3\label{#2}\end{lem}
\begin{proof}#4\end{proof}}
{\begin{lem}[#1]#3\label{#2}\end{lem}
\begin{proof}#4\end{proof}}
}
\newcommand{\nprlemma}[3][]{
\ifthenelse{\equal{#1}{}} {\begin{lem}#3\label{#2}\end{lem}}
{\begin{lem}[#1]#3\label{#2}\end{lem}} }

\newtheorem{cor}{Corollary}[section]
\newcommand{\corollary}[4][]{
\ifthenelse{\equal{#1}{}} {\begin{cor}#3\label{#2}\end{cor}
\begin{proof}#4\end{proof}}
{\begin{cor}[#1]#3\label{#2}\end{cor}
\begin{proof}#4\end{proof}}
}
\newcommand{\nprcorollary}[3][]{
\ifthenelse{\equal{#1}{}} {\begin{cor}#3\label{#2}\end{cor}}
{\begin{cor}[#1]#3\label{#2}\end{cor}} }

\newcommand{\nref}[1]{(\ref{#1})}
\newcommand{\labeq}[2]{
\begin{equation}#2\label{#1}\end{equation}
}

\newcommand{\suchthat}{;}
\newcommand{\set}[1]{\left\{#1\right\}}

\newcommand{\Sp}[4]     
{ #1^{#2}_{#3}(#4) }

\newcommand{\Lp}[2]     %
{\Sp{L}{#1}{}{#2}}

\newcommand{\Lii}[1]        %
{\Sp{L}{2}{}{#1}}

\newcommand{\Hi}[1]        %
{\Sp{H}{1}{}{#1}}

\newcommand{\Hii}[1]        %
{\Sp{H}{2}{}{#1}}

\newcommand{\dblSp}[4]     %
{\Sp{#1}{#2}{}{#3 ; #4}}

\newcommand{\dblLp}[3]
{\dblSp{L}{#1}{#2}{#3}}   %

\newcommand{\dblLii}[2]     %
{\dblLp{2}{#1}{#2}}

\newcommand{\C}[2][]     %
{\Sp{C}{#1}{}{#2}}

\newcommand{\dblC}[3][]
{\dblSp{C}{#1}{#2}{#3}}   %
\newcommand{\dtd}[1]{\dot{#1}}      

\newcommand{\ddtd}[1]            
{\stackrel{\ndot\ndot}{#1}}

\newcommand{\pd}[2]{\frac{\partial #1}{\partial #2}}

\newcommand{\spd}[2]{\frac{\partial^{2} #1}{{\partial #2}^{2}}} 

\newcommand{\ptd}[2][t]{\pd{#2}{#1}} 

\newcommand{\grad}[2][]{
\ifthenelse{\equal{#1}{}}{\nabla#2}{\nabla\left(#2\right)} }

\newcommand{\dvrg}[2][]{
\ifthenelse{\equal{#1}{}}{\nabla \! \cdot \! #2}{\nabla \! \cdot\!
\left(#2\right)} }

\newcommand{\newint}[4]{\int_{#1}^{#2}#3\,{\rm d}#4}

\newcommand{\domint}[1]         
{\newint{\dom}{}{#1}{x}}

\newcommand{\eval}[3][]
{\ifthenelse{\equal{#1}{}} {\left.#2\right|_{#3}}
{\left.#2\right|_{#1=#3}}}

\newcommand{\infint}[2]         
{\newint{0}{\infty}{#1}{#2}}
\newcommand{\sinfint}[1]         
{\infint{#1}{s}}
\newcommand{\minusplusinfint}[2]         
{\newint{-\infty}{\infty}{#1}{#2}}
\newcommand{\sminusplusinfint}[1]         
{\minusplusinfint{#1}{s}}
\newcommand{\cyclint}[2]        
{\oint_{0}^{#1}#2\,{\rm d}t}

\newcommand{\modulo}[1]{\left\vert#1\right\vert}
\newcommand{\sqmod}[1]{\left\vert#1\right\vert^{2}}

\newcommand{\recipr}[1]{\frac{1}{#1}}
\newcommand{\hist}[2][t]{#2^{#1}}

\newcommand{\histfunct}[1][]{\ifthenelse{\equal{#1}{}}{\varphi_{t}^{*}}{\var
phi_{t}^{\mathcal{#1}}}}
\newcommand{\inhistfunct}[1][]{\ifthenelse{\equal{#1}{}}{\varphi_{0}^{*}}{\v
arphi_{0}^{\mathcal{#1}}}}

\newcommand{\adj}[1]{\widetilde{#1}}

\newcommand{\Dom}[1]{\mathcal{D}(#1)}

\newcommand{\Lapl}[1]{\hat{#1}}
\newcommand{\funct}[3]{#1 : #2 \longrightarrow #3}

\newcommand{\R}{\mathbb{R}}
\newcommand{\Rplus}{\mathbb{R}^{+}}
\newcommand{\Compl}{\mathrm{C}}

\newcommand{\real}[1]{\Re\left\{#1\right\}}

\newcommand{\bv}{\gamma_0}
\newcommand{\bvk}{\lambda}
\newcommand{\Op}{A}
\newcommand{\adjOp}{\adj{A}}
\newcommand{\state}[1][]{\ifthenelse{\equal{#1}{}} {\sigma} {\sigma_{#1}} }
\newcommand{\adjstate}[1][]{
\ifthenelse{\equal{#1}{}} {\adj{\sigma}} {\adj{\sigma}_{#1}} }
\newcommand{\statetd}[1][]{
\ifthenelse{\equal{#1}{}} {\dtd{\sigma}} {\dtd{\sigma}_{#1}} }

\newcommand{\statesp}{\mathcal{K}}

\newcommand{\freen}[2][]{
\ifthenelse{\equal{#1}{}} {\psi^{#2}} {\psi^{#2}_{#1}} }
\newcommand{\freentd}[2][]{
\ifthenelse{\equal{#1}{}} {\dtd{\psi}^{#2}} {\dtd{\psi}^{#2}_{#1}} }
\newcommand{\intfreen}[1][]{\freen[\dom]{#1}}

\newcommand{\fabrizio}{\psi}

\newcommand{\dom}{\Omega}

\newcommand{\ndot}{\cdot}       

\newcommand{\half}[1][]{\ifthenelse{\equal{#1}{}}{\frac{1}{2}}{\frac{#1}{2}}
}
\newcommand{\ahist}[1][t]{\hist[#1]{\breve{a}}}

\newcommand{\adjahist}[1][t]{\hist[#1]{\adj{a}}}

\newcommand{\Lfnct}[2][t_{0}]{\mathcal{L}_{#1}\left(#2\right)}

\newcommand{\ray}[1][]{\ifthenelse{\equal{#1}{}}{l}{l(#1)}}
\newcommand{\lapldom}[1][\expdecconstc]{\mathcal{D}_{#1}}

\newcommand{\bndkernelconst}{k_0}
\newcommand{\expdecconsta}{c_{1}}
\newcommand{\expdecconstb}{c_{2}}

\journal{Journal of Mathematical Analysis and Applications}

\begin{document}

\begin{frontmatter}

\title{On the exponential decay of the Euler-Bernoulli beam with boundary energy dissipation\tnoteref{label1}}
\tnotetext[label1]{Research performed under the
auspices of G.N.F.M. - I.N.d.A.M. and partially supported by Italian
M.I.U.R..}
\author{Barbara Lazzari \corref{corresponding}}
\cortext[corresponding]{corresponding author: phone +390512094495, fax  +390512094490.}
\ead{lazzari@dm.unibo.it}
\author{Roberta Nibbi \corref{}}
\ead{nibbi@dm.unibo.it}
\address{Dipartimento di Matematica,
Alma Mater Studiorum - Universit\`{a} di Bologna, Piazza di Porta S. Donato 5,
40126 Bologna, Italy}

\begin{abstract}
We study the asymptotic behavior of the Euler-Bernoulli beam which is clamped at one end and free at the other end. We apply a boundary control with memory at the free end of the beam and prove that the ``exponential decay'' of the memory kernel is a necessary and sufficient condition for the exponential decay of the energy.

\end{abstract}


\end{frontmatter}

\section {Introduction}

In this paper we study the long time behavior of the Euler-Bernoulli beam clamped at one end and free at the other end (cantilever beam).
Here, for simplicity and without loss of generality, we assume the length, the density and the flexural rigidity  of the beam equal to $1$.

The dynamic problem in $(0,1) \times \Rplus$ is therefore described by the well known equation of motion
\labeq{eq1}{u_{tt}(x,t)+ u_{xxxx}(x,t)= 0,}
together with the boundary conditions
\labeq{bnd1}{u(0,t) = u_x(0,t) = 0}
and
$${u_{xx}(1,t)= \beta(t), \qquad u_{xxx}(1,t) = \Gamma(t),}$$
where $\beta(t)$ and $\Gamma(t)$ are boundary control terms applied to the free end of the beam.

The boundary feedback stabilization problem of this model, that is the problem of finding boundary controls capable to guarantee the exponential stability, has been studied at length (see \cite{Chen1987, GuoWang2005, GuoYu2001, GuoHuang2004, Krall1989} and references therein).

In this paper we choose the following boundary control with memory
\labeq{introboundcond}{ u_{xx}(1,t) = 0, \qquad  u_{xxx}(1,t) = \bv u_t(1,t) +
\int_0^{\infty}{\bvk(s)u_t(1,t-s) \, {\rm d}s} ,} where $\bv \in \Rplus$
and the
memory kernel  $\funct{\bvk}{\Rplus}{\R}$ belongs to
$\Lp{1}{\Rplus} \cap \Hii{\Rplus}$.

This control has been already proposed in \cite{Park-Kim2008} and \cite{Park-Kim2005} where, in presence of further structural dampings, it has been proved that the energy has the same rate of decay (exponential or polynomial) of the memory kernel.

Here, generalizing energy estimates obtained in \cite{Chen1987} for the boundary control
\labeq{precboundcond}{u_{xx}(1,t) = 0, \qquad  u_{xxx}(1,t) = \bv u_t(1,t), \quad \bv > 0}
we prove that, whenever the memory kernel decays exponentially, so does the energy of the system.

It is  interesting to observe that boundary condition \nref{precboundcond} ensures the exponential decay of the energy for the cantilever beam, while, in the presence of a memory term at the boundary, such a result is not assured. Indeed, we shall show that the condition
\labeq{expdeccond}{\int_0^{\infty} {e^{\delta t}\bvk(t)} \, {\rm d}t < \infty}
for some $\delta > 0$, turns out to be necessary for the exponential decay of the solution.

Finally, we observe that this control can be also seen as a generalization of the case of a mass attached to the free end of the beam \cite{ConradMorgu1998, Li2001}. If, in fact, we choose an exponential function as memory kernel, by differentiating \nref{introboundcond} with respect to time, we obtain
$$u_{xxx}(1,t) - m u_{tt}(1,t) = \alpha u_t(1,t) - \beta u_{xxxt}(1,t).$$

The outline of the paper is the following.

In Section \ref{Wellpos} we prove existence, uniqueness and
regularity of solutions for the related initial boundary problem via semigroup theory.
In Section \ref{Expdec}, after developing the needed estimates,
we prove that the exponential decay of the memory kernel turns out to be a necessary and sufficient condition for the exponential decay of the energy.

\setcounter{equation}{0}

\section{Well posedness}\label{Wellpos}

Let us consider problem \nref{eq1}--\nref{introboundcond} together with the initial conditions
\labeq{initcond}{
u(x,0) = u_{0}(x), \quad u_t(x,0) = v_{0}(x), \qquad x \in (0,1).
}
From now on, whenever no ambiguity arises, we shall drop the $x$ variable  and we shall refer to problem
\nref{eq1}--\nref{introboundcond}-\nref{initcond} as to \emph{problem $P$}.

The aim of this section is the proof of the well-posedness of \emph{problem $P$} via semigroup theory.
To this end  we introduce the \textit{past history} $${w}(1,t-s) =u(1,t-s) - u(1,t)$$ and rewrite the boundary control term in \nref{introboundcond} as follows
\labeq{boundcond}{u_{xxx}(1,t) = \bv u_t(1,t) +
\int_0^{\infty}\bvk'(s){w}(1,t-s) \, {\rm d}s.}
Evolution problems presenting boundary controls of  memory type similar to (\ref{boundcond}) have been studied  in several fields, making use of  the concepts of dissipative boundary  and boundary energy (see \cite{BoselloLazzariNibbi2007} and references therein), and the related solutions have been  usually found among those with ``finite energy". 
It should however be noted that in presence of memory, the energy-type  functional turns out to be non-unique.

Following this approach, (\ref{boundcond})  is compatible with the definition of dissipative boundary if the memory kernel satisfies 
\labeq{fourier}{\omega \int_0^{\infty} \bvk^{'}(s) \sin(\omega s) \, {\rm d} s < 0, \qquad \omega \neq 0.}
If (\ref{fourier}) holds, reasoning as in  \cite{BoselloLazzariNibbi2007}, it is possible  to prove the well-posedness of \emph{problem $P$}   in the past histories space ${\cal{H}}_1$, for which the energy functional 
\labeq{maximal}{
\half \int_0^{\infty}\int_0^{\infty} \frac{\partial^2 \lambda(|s_1 - s_2|)}{\partial s_1 \partial s_2} {w}(1,t-s_1){w}(1,t-s_2) \, {\rm d}s_1  \, {\rm d}s_2}
 is finite. 
 On the other hand
 this result is not satisfactory, because ${\cal{H}}_1$ does not contain even all the bounded histories (indeed \cite{FGM1994} sinusoidal histories do not belong to ${\cal{H}}_1$) and it is therefore desirable to obtain well posedness results in wider  past histories spaces.

Certainly the space  ${\cal {H}}_2$ of the past histories  for which 
\labeq{graffi}{
- \half \int_0^{\infty} \bvk^{'}(s) |{w}(1,t-s)|^2 \, {\rm d}s<\infty
}
contains at least all the bounded histories.
However, to obtain  well-posedness results  in this space the memory kernel must satisfy the following more restrictive hypotheses
\labeq{bvkhyp}{\bvk^{'}(s) < 0, \quad \bvk^{''}(s) \geq 0, \qquad
 s \in \Rplus.}

As observed in \cite{DFG2006}, it is not necessary to know the past history $w$ at all times, because two different histories $w_1$ and $w_2$ satisfying
$$
\int_0^{\infty} \bvk^{'}(\tau + s){w}_1(1,t-\tau) \, {\rm d}\tau =\int_0^{\infty} \bvk^{'}(\tau + s){w}_2(1,t-\tau) \, {\rm d}\tau\quad  \forall s \in \Rplus \,,
$$
lead to the same boundary control term in (\ref{boundcond}).
It is therefore convenient a formulation which relies only on the minimal information required to determine the boundary control. 
To this end, here we study \emph{problem $P$}  in terms of the new variable
\labeq{ahistdef}{\ahist(1,s)
= - \int_0^{\infty} \bvk^{'}(\tau + s){w}(1,t-\tau) \, {\rm d}\tau , \qquad s \in \Rplus }
so that
 \nref{boundcond} takes the form
\labeq{ahistbndcnd}{u_{xxx}(1,t) - \bv u_t(1,t) = - \ahist(1,0)\,.} 
By introducing the functional 
\labeq{benergy}{ \psi_b(t) =
-\half
\int_0^{\infty} \recipr{\bvk^{'}(s)}  \modulo{\frac{\partial \ahist(1,s)}{\partial s}}^{2}  \, {\rm d}s.}
 we are able to achieve well posedness results in a space wider than  ${\cal {H}}_2$,  leaving unchanged the hypotheses (\ref{bvkhyp}) on the kernel. In fact the following proposition holds.

\proposition{GraffiandFabrizioestim}{Let $w\in {\cal {H}}_2$ and $\ahist$ defined in terms of $w$  through { \rm(\ref{ahistdef})}, then the functional  { \rm(\ref{benergy})} is finite.} 
{Rewriting  { \rm(\ref{benergy})} in terms of $w$ and using classical inequalities, it follows that
$$
 \psi_b(t)= -\half \int_0^\infty \recipr{\bvk^{'}(s)}  \left|\int_0^{\infty} \bvk^{''}(\tau + s)w(1,t-\tau) \, {\rm d}\tau\right|^{2}  \, {\rm d}s\qquad\qquad\qquad\qquad\qquad
$$
  $$\qquad \leq
-\half \int_0^\infty \frac{ \int_0^\infty \bvk^{''}(\tau +s)\, {\rm d}\tau }{\bvk^{'}(s)}   \left( \int_0^\infty \bvk^{''}(\tau +s)\left|{w}(1,t-\tau)\right|^2\,
{\rm d}\tau \right) \,{\rm d}s \qquad\qquad$$
  $$\qquad =
\half \int_0^\infty  \int_0^\infty \bvk^{''}(\tau +s)\left|{w}(1,t-\tau)\right|^2\,
{\rm d}\tau \,{\rm d}s =-
\half \int_0^\infty \bvk^{'}(\tau )\left|{w}(1,t-\tau)\right|^2\,
{\rm d}\tau\,. $$
The thesis follows immediately from the hypothesis $w\in {\cal {H}}_2$.}

Moreover, the functional  $\psi_b$ satisfies
 \labeq{bndfabrtder} { \dot{\psi}_b(t) = - \ahist(1,0)u_t(1,t)
- \half
\int_0^{\infty} \frac{\bvk^{''}(s)}{\left[\bvk^{'}(s)\right]^{2}}
\modulo{\frac{\partial \ahist(1,s)}{\partial s}}^{2}
 \, {\rm d}s +  \half
\recipr{\bvk^{'}(0)}\modulo{\frac{\partial \ahist(1,0)}{\partial s}}^{2},
}
since the derivative of $\ahist$ with respect to $t$ is given by
$$ \ptd[s]{}\ahist(1,s)
-\bvk(s)v(1,t).
$$
Finally we observe that
$$ |\ahist(1,s)|^2 = \left| -
\int_0^{\infty} \frac{\partial \ahist(1,s+\tau)}{\partial \tau} \, {\rm d} \tau
\right|^2 \leq
$$ \labeq{correz}{
\leq \int_0^{\infty} - \lambda^\prime(s+\tau) \, {\rm d} \tau
\int_0^{\infty} -\frac{1}{\lambda^\prime(s+\tau)} \modulo{\frac{\partial
\ahist(1,s+\tau)}{\partial \tau}}^{2} \,{\rm d} \tau \leq 2 \lambda(s)
\psi_b(t);}
in particular   \nref{ahistbndcnd} and \nref{correz} yield \labeq{continuity}{  |u_{xxx}(1,t)|^2  \leq 2
\gamma_0^2 |u_t(1,t)|^2 + 4 \lambda(0) \psi_b(t). }

Let us now define $v = u_t$ and introduce the state $\state = (v,u_{xx},\ahist)$
and the total energy
\labeq{totalenergy}{ \psi(t) =
\underbrace{\half\int_0^1 \left[|v(t)|^{2}+ |u_{xx}(t)|^{2} \right] {\rm d}x}_{\intfreen(t)} \; \underbrace{-\half
\int_0^{\infty} \recipr{\bvk^{'}(s)}  \modulo{\frac{\partial \ahist(1,s)}{\partial s}}^{2}  \, {\rm d}s}_{\psi_b(t)}.}
This energy coincides with the energy proposed in \cite{ConradMorgu1998} when the memory kernel is an exponential function. 
 
 We rewrite the \emph{problem $P$} as an abstract
first-order Cauchy problem as follows: \labeq{abstrprobl}{ \left\{
\begin{array}{rcl}
\dtd{\state}(t)&=& \Op \state(t)
\\
\state(0)&=&\state[0]
\end{array}
\right. } with $\state[0] =
(v_{0},{u_0}_{xx},\breve{a}^0(1,\cdot))$ and

\labeq{op}{ \Op \state(t) = \left( - u_{xxxx}(t), {v}_{xx}(t), \ptd[s]{\ahist(1,s)}
-\bvk(s)v(1,t) \right). } As said before, the natural setting in which to
look for existence and uniqueness of solutions for problem
\nref{abstrprobl} is the \emph{admissible states space} $\statesp$,
consisting in those states $\state$ for which the total energy
\nref{totalenergy} is finite.
We endow $\statesp$ with the inner product \[ \langle
\state[1](t),\state[2](t) \rangle = \int_0^1 \left[
v_{1}(t) v_{2}(t) + u_{1xx}(t) u_{2xx}(t) \right] {\rm d}x  -
\int_0^{\infty}\recipr{\bvk^{'}(s)}\pd{\ahist_{1}(1,s)}{s}\pd{\ahist_{2}
(1,s)}{s} \,{\rm d} s , \] where  $\state[i](t) = (v_i(t),{u_i}_{xx}(t), \ahist_{i}(1,\cdot))$, for $i=1,2$, so that
\[
\langle \state(t), \state(t) \rangle = \| \state(t)\|^{2} = 2 \psi(t).
\]
We denote by $\Dom{\Op}$ the domain of the operator $\Op$, namely
\[
\Dom{\Op}= \set{\state \in \statesp \suchthat \Op\state \in
\statesp \mbox{ and boundary conditions \nref{bnd1}-\nref{ahistbndcnd} hold}}
\]
and claim that the operator $\Op$ is dissipative.

In fact if $\state \in \Dom{\Op}$, we have
\[
\langle \Op \state(t),\state(t) \rangle =
\int_0^1{ \left[
- u_{xxxx}(t)v(t) + u_{xx}(t)v_{xx}(t)\right]} \, {\rm d}x +\]
\[
 - \int_0^{\infty} \recipr{\bvk^{'}(s)}\frac{\partial}{\partial s} \left[\frac{\partial \ahist(1,s)}{\partial s}
- \bvk(s)v(1,t)\right] \frac{\partial \ahist(1,s)}{\partial s} \, {\rm d}s =
\]
\[
 = -u_{xxx}(1,t) v(1,t) -  \int_0^{\infty}
\recipr{\bvk^{'}(s)} \spd{\ahist(1,s)}{s} \pd{\ahist(1,s)}{s} \, {\rm d}  s
- v(1,t)\ahist(1,0)=
\]
\[= - \bv \modulo{v(1,t)}^{2}
- \half
\int_0^{\infty}\frac{\bvk^{''}(s)}{\left[\bvk^{'}(s)\right]^{2}}
\modulo{\frac{\partial \ahist(1,s)}{\partial s}}^{2}  \,{\rm d}s +
 \half
\recipr{\bvk^{'}(0)}\modulo{\frac{\partial \ahist(1,0)}{\partial s}}^{2}
\leq 0.
\]
We now proceed to show that also $\adjOp$, the adjoint of $\Op$, is
dissipative so that, thanks to the Lumer-Phillips theorem \cite{Pazy}, $\Op$
generates a $C_0$-semigroup.

Let $\adjstate = (\adj{v},\adj{u}_{xx}, \adjahist)$ be in $\statesp$ and consider the boundary
conditions \labeq{adjboundcond}{\adj{u}(0,t) = \adj{u}_x(0,t) =\adj{u}_{xx}(1,t)= 0, \quad  \adj{u}_{xxx}(1,t)= - \bv \adj{v}(1,t) - \adjahist(1,0).} Denoting by $H$
the Heaviside function and introducing $j[\adjahist(1, \cdot)]$
such that
\[
\frac{\partial }{\partial s}j[\adjahist(1, \cdot)](s) =
-\bvk^{'}(s)\frac{\partial}{\partial s} \left( \frac{H(s)}{\bvk^{'}(s)} \right) \frac{\partial}{\partial s}\adjahist(1,s),
\]
we claim that
$$\adjOp\adjstate (t) = \left( \adj{u}_{xxxx}(t),- \adj{v}_{xx}(t),
- \frac{\partial}{\partial s}\adjahist(1,s)+\bvk(s)\adj{v}(1,t) + j[\adjahist(1, \cdot)](s)\right)$$ and
that the domain of $\adjOp$ is
\[
\Dom{\adjOp} = \set{ \adjstate \in \statesp \suchthat
\adjOp\adjstate \in \statesp \; \mbox{and the boundary conditions}
\; \nref{adjboundcond} \;
 \mbox{hold}}.
\]
Let us now compute $\langle \Op\state,\adjstate \rangle$, where
$\state \in \Dom{\Op}$ and  $\adjstate \in \Dom{\adjOp}$:
\[
\langle \Op\state(t), \adjstate(t) \rangle =
\int_0^1 \left[ - u_{xxxx}(t) \adj{v}(t) +
v_{xx}(t) \adj{u}_{xx}(t)
  \right] \, {\rm d} x +
\]
 \[ -
\int_0^{\infty} \recipr{\bvk^{'}(s)}\frac{\partial}{\partial s}\left(
\frac{\partial \ahist(1,s)}{\partial s}-\bvk(s)v(1,t) \right) \frac{\partial \adjahist(1,s)}{\partial s} \, {\rm d} s
=
\]
\[ = \int_0^1 \left[v(t)\adj{u}_{xxxx}(t) - \adj{v}_{xx}(t) u_{xx}(t) \right] \,{\rm d} x +
\]
\[
- v(1,t) \left[ \adj{u}_{xxx}(1,t) + \bv \adj{v}(1,t) + \adjahist(1,0)\right] +\]
\[+
\int_0^{\infty}
\recipr{\bvk^{'}(s)}\frac{\partial \ahist(1,s)}{\partial s}\frac{\partial}{\partial s} \left(
\frac{\partial \adjahist(1,s)}{\partial s} -\bvk(s)\adj{v}(1,t)\right)\, {\rm d} s  +\]
\[+
\int_0^{\infty} \frac{\partial}{\partial s} \left(\recipr{\bvk^{'}(s)}\right)\frac{\partial \ahist(1,s)}{\partial s}\frac{\partial \adjahist(1,s)}{\partial s} \, {\rm d} s +
\recipr{\bvk^{'}(0)}\frac{\partial \ahist(1,0)}{\partial s} \frac{\partial \adjahist(1,0)}{\partial s},
\]
so that, if $\adjstate$ satisfies the boundary conditions
\nref{adjboundcond}, we have
\[
\langle \Op\state(t), \adjstate(t) \rangle =  \langle \state(t),-\Op\adjstate(t)
\rangle  + \int_0^{\infty} \frac{\partial}{\partial s} \left(\frac{H(s)}{\bvk^{'}(s)}\right)
\frac{\partial \ahist(1,s)}{\partial s}\frac{\partial \adjahist(1,s)}{\partial s}\, {\rm d} s= \langle \state(t), \adjOp\adjstate(t)\rangle.
\]
Now observe that, for $\adjstate \in \Dom{\adjOp}$, we have
\[
\langle \adjOp\adjstate(t), \adjstate(t)\rangle = -
\bv\modulo{\adj{v}(1,t)}^{2}
-\half \int_0^{\infty} \frac{\bvk^{''}(s)}{\left(\bvk^{'}(s)\right)^{2}}
\modulo{\frac{\partial \adjahist(1,s)}{\partial s}}^{2} \, {\rm d} s
+ \half \recipr{\bvk^{'}(0)}\modulo{\frac{\partial \adjahist(1,0)}{\partial s}}^{2}
\leq 0.
\]
Finally, making use of well known results on the semigroup theory \cite{DS}, it is
possible to state the following theorem establishing the well
posedness of \emph{problem $P$}:
\nprtheorem{wellpos}{If $\state[0] \in \Dom{\Op}$, then problem {\rm \nref{abstrprobl}}
admits one and only one strict solution $\state \in
\dblC[1]{\Rplus}{\statesp} \cap \dblC{\Rplus}{\Dom{\Op}}$. }

\setcounter{equation}{0}
\section{Exponential decay}\label{Expdec}

In order to show that an exponential decay of the energy
\nref{totalenergy} occurs over time, it is necessary to impose
further conditions. More precisely, we
shall assume that $\bv > 0$ and that there exists $\bndkernelconst > 0$ such that
\labeq{H1}{\bvk''(s) + \bndkernelconst\bvk'(s) \geq 0,  \quad s \in \Rplus.}
It should be remarked that
\nref{H1} is in some sense a hypothesis of exponential decay
on the memory kernel $\lambda$, in the sense that it easily yields
\[
| \bvk'(s) | = - \bvk'(s) \le c_0 e^{- \kappa s},
\qquad  s \in \Rplus
\]
for a suitable positive constant $c_0$.

The main result of this section is the following
\theorem{Expdecay}{Let $\state$ be a solution of
{\rm\nref{abstrprobl}}.
If $\gamma_0 >0$ and the memory kernel satisfies \nref{bvkhyp} and \nref{H1}, then
there exist two positive constants $\expdecconsta$ and
$\expdecconstb$ such that
\[\fabrizio(t) \leq \expdecconstb e^{-\expdecconsta t}\fabrizio(0).\]}
{Thanks to the semigroup properties proved in the preceding section, in order to obtain the exponential decay of the total energy it is sufficient to show that (see, for instance, Th. 4.1 in \cite{Pazy})
\labeq{L_4}{\psi(t) \leq
 \frac{h_1}{(t + h_2)}.}
To this aim we introduce the functional
\[ \Lfnct{t}
=  (t+t_{0})\psi(t) + \int_0^1
x u_t(t)u_x(t)\,{\rm d} x
\]
and prove that, for $t_{0}$ sufficiently large, it is monotonically non-increasing for every
solution of \nref{abstrprobl}.

In fact, if $\state$ is
a solution of \nref{abstrprobl}, then it is easy to show that
\[
\dot{\psi}(t)= - \bv |u_t(1,t)|^2 + \half
\recipr{\bvk^{'}(0)}\modulo{\frac{\partial \ahist(1,0)}{\partial s}}^{2} - \half
\int_0^{\infty} \frac{\bvk^{''}(s)}{\left[\bvk^{'}(s)\right]^{2}}
\modulo{\frac{\partial \ahist(1,s)}{\partial s}}^{2}
 \, {\rm d}s  \]
and \nref{H1} yields
\labeq{L_1}{
\dot{\psi}(t) \leq  - \bv |u_t(1,t)|^2 - \bndkernelconst \psi_b(t).}
Moreover,
\[
\frac{d}{dt} \left( \int_0^1
x u_t(t)u_x(t)\,{\rm d} x \right) = - \psi_{\dom}(t) - \int_0^1
|u_{xx}(t)|^2 \,{\rm d} x + \frac{1}{2}|u_t(1,t)|^2 -  u_{x}(1,t) u_{xxx}(1,t) \leq
\]
\[
\leq - \psi_{\dom}(t) - \int_0^1
|u_{xx}(t)|^2 \,{\rm d} x + \frac{1}{2}|u_t(1,t)|^2 + |u_{x}(1,t)|^2 + \frac{1}{4}|u_{xxx}(1,t)|^2.
\]
Recalling the boundary conditions \nref{ahistbndcnd} and the inequality \nref{continuity}, we get
\labeq{L_2}{
\frac{d}{dt} \left( \int_0^1
x u_t(t)u_x(t)\,{\rm d} x \right)
\leq - \psi_{\dom}(t) + \frac{1}{2}|u_t(1,t)|^2 + \frac{1}{2} \left[
\gamma_0^2 |u_t(1,t)|^2 + 2 \lambda(0) \psi_b(t)\right].
}
Finally, thanks to \nref{L_1} and \nref{L_2},
\labeq{L_3}{\frac{d}{dt} \Lfnct{t} = (t+t_{0})\dot{\psi}(t) + \psi(t) + \frac{d}{dt}\left(\int_0^1
x u_t(t)u_x(t)\,{\rm d} x \right)
\leq
}
\[
\leq \left[\frac{1}{2} \left( 1 +  \bv^2\right) - \bv (t+ t_0)\right]|u_t(1,t)|^2 + \left[ \lambda(0) - \bndkernelconst(t+ t_0)\right]\psi_b(t).
\]
On the other hand, using the classical inequalities of Cauchy-Schwarz and Poincar\'{e},  it is easy to prove that
\[ \modulo{\int_0^1
x u_t(t)u_x(t)\,{\rm d} x } \leq  \psi_{\Omega}(t) \leq  \psi(t),
\]
so that
\labeq{L_5}{\Lfnct{t} - \Lfnct{0} \geq (t+t_0 - 1) \psi(T) - (t_0 + 1)\psi(0).
}
Finally, choosing
\[
t_0 \geq \max \left\{\frac{1}{2 \bv} \left( 1 +  \bv^2\right), \frac{ \lambda(0)}{\bndkernelconst}, 1 \right\},
\]
it follows that
\[0 \geq
 \Lfnct{t} - \Lfnct{0} \geq (t + t_0 - 1) \psi(t) - (t_0 + 1) \psi(0)
\]
or, equivalently,
\[\psi(t) \leq
 \frac{(t_0 + 1)}{(t + t_0 - 1)} \psi(0),
\]
which coincides with \nref{L_4} by putting $h_1 = (t_0 + 1) \psi(0)$ and $h_2= (t_0 - 1)$.
}

The previous theorem guarantees the exponential decay not only of the internal energy of the beam but also of $\psi_b$. Therefore, thanks to estimate \nref{correz}, also $\ahist(1,0)$ decays exponentially but we cannot conclude the same for $u_t(1, \cdot)$ and $u_{xxx}(1, \cdot)$ separately.
However, as already noted in \cite{ConradMorgu1998}, if the initial data are sufficiently smooth (for example $\state[0]\in \Dom{\Op}$) we obtain also the exponential decay of $u_t(1, \cdot)$ and $u_{xxx}(1, \cdot)$.

We close this section showing that the ``exponential decay '' of the
control memory kernel turns out to be a necessary condition for the
``exponential decay'' of the solution of \emph{problem $P$}. To be more precise, we give the following definition.
\definition{expdecdef}{\rm{A function $u$ \emph{decays exponentially} if there exists a
positive constant $\delta$ such that}
\[\infint{e^{\delta t}\modulo{u(t)}}{t} < \infty.\]}
We will obtain the exponential decay of the memory kernel as a consequence of the following result (see \cite{Mu}, Theorem 2):
\nprlemma{Mulemma}{A non-negative function $\lambda \in \Lp{1}{\Rplus}$ decays exponentially if there exist a neighborhood $U \subset \mathbb{C}$ of $0$  and
a holomorphic function $\funct{f}{U}{\mathbb{C}}$ such that the Laplace transform of $\lambda$ coincides with $f$ in $U \cap \mathbb{C}^+$, where $\mathbb{C}^+ = \set{z \in \mathbb{C} \, ; \,\real{z} \geq 0}$. }

\nprtheorem{expdec}{Let $u$ be a solution
of problem $P$ with null sources and such that
\labeq{expdechyp}{\infint{e^{2\delta
t}\left(\sqmod{u_t(1,t)} + \sqmod{{u}_{xxx}(1,t)}\right)}{t} <
\infty} for some $\delta> 0$,  then $\bvk$
decays exponentially.}
\noproppr[expdec]{First of all we observe that if $u$ is a solution of \emph{problem $P$} satisfying \nref{expdechyp} then the Laplace transforms
$$\Lapl{u}_t(1,z)  =  \infint{e^{-zt}u_t(1,t)}{t}, \quad \Lapl{u}_{xxx}(1,z)  =  \infint{e^{-zt}u_{xxx}(1,t)}{t}$$
are holomorphic functions in $ \lapldom
= \set{z \in \Compl \suchthat \real{z} > - \delta}$ and, since $\lambda \in \Lp{1}{\Rplus}$, the Laplace transform of the boundary condition \nref{introboundcond} 
\labeq{boundcondtransf}{{\Lapl{u}_{xxx}}(1,z)
= \left(\bv + \Lapl{\bvk}(z)\right)\Lapl{u}_t(1,z)}
is well defined for $z \in \mathbb{C}^+$.

Reasoning as in the proof of Theorem \ref{Expdecay}, it is easy to show that the (constant in time) energy $\psi_{\Omega}$ , associated to the cantilever beam problem \nref{eq1}-\nref{bnd1} with a vanishing boundary control term $\Gamma(t)$ at the free end,   satisfies
\[
(t -2) \psi_{\Omega}(0)\leq \int_0^t |u_t(1, \tau)|^2 \,{\rm d} \tau ;
\]
therefore, if we give the additional boundary condition
$u_t(1,t) \stackrel{t}{\equiv} 0$, the cantilever beam problem admits only the trivial solution.

Consequently, if $u$ is a non-trivial
solution of \emph{problem $P$} satisfying \nref{expdechyp},
it exists a non-negative integer $k$  such that
\labeq{nonzeroder}{\frac{\partial^k}{\partial z^k} \Lapl{u}_t(1,0) \neq 0.}
If $k_0$ is the first integer for which \nref{nonzeroder} holds, then
 \labeq{g}{\Lapl{u}_t(1,z)  = z^{k_0-1}g(z)}
with $g$ holomorphic on $\lapldom$ and $g(0) \neq 0$.

Similarly, by virtue of \nref{boundcondtransf}, there exists G, holomorphic in $\lapldom$, such that
 \labeq{G}{\Lapl{u}_{xxx}(1,z)  = z^{k_0-1}G(z).}
Therefore, we conclude that
 \[\Lapl{\bvk}(z) = \frac{G(z)}{g(z)} -\bv, \qquad z \in \mathbb{C}^+\]
where the right-hand side is a holomorphic function in  a neighborhood
$\mathcal{U}$ of $0$, since  $g(0) \neq 0$.}


\begin{thebibliography}{15}
\expandafter\ifx\csname natexlab\endcsname\relax\def\natexlab#1{#1}\fi
\providecommand{\bibinfo}[2]{#2}
\ifx\xfnm\relax \def\xfnm[#1]{\unskip,\space#1}\fi
\bibitem[{Chen et~al.(1987)Chen, Krantz, Ma, Wayne, and West}]{Chen1987}
\bibinfo{author}{G.~Chen}, \bibinfo{author}{S.~G. Krantz},
  \bibinfo{author}{D.~W. Ma}, \bibinfo{author}{C.~E. Wayne},
  \bibinfo{author}{H.~H. West},
\newblock \bibinfo{title}{The {E}uler-{B}ernoulli beam equation with boundary
  energy dissipation},
\newblock in: \bibinfo{booktitle}{Operator methods for optimal control problems
  ({N}ew {O}rleans, {L}a., 1986)}, volume \bibinfo{volume}{108} of
  \textit{\bibinfo{series}{Lecture Notes in Pure and Appl. Math.}},
  \bibinfo{publisher}{Dekker}, \bibinfo{address}{New York},
  \bibinfo{year}{1987}, pp. \bibinfo{pages}{67--96}.
\bibitem[{Guo et~al.(2005)Guo, Wang, and Yung}]{GuoWang2005}
\bibinfo{author}{B.-Z. Guo}, \bibinfo{author}{J.-m. Wang},
  \bibinfo{author}{S.-P. Yung},
\newblock \bibinfo{title}{On the {$C\sb 0$}-semigroup generation and
  exponential stability resulting from a shear force feedback on a rotating
  beam},
\newblock \bibinfo{journal}{Systems Control Lett.} \bibinfo{volume}{54}
  (\bibinfo{year}{2005}) \bibinfo{pages}{557--574}.
\bibitem[{Guo and Yu(2001)}]{GuoYu2001}
\bibinfo{author}{B.-Z. Guo}, \bibinfo{author}{R.~Yu},
\newblock \bibinfo{title}{The {R}iesz basis property of discrete operators and
  application to a {E}uler-{B}ernoulli beam equation with boundary linear
  feedback control},
\newblock \bibinfo{journal}{IMA J. Math. Control Inform.} \bibinfo{volume}{18}
  (\bibinfo{year}{2001}) \bibinfo{pages}{241--251}.
\bibitem[{Guo and Huang(2004)}]{GuoHuang2004}
\bibinfo{author}{F.~Guo}, \bibinfo{author}{F.~Huang},
\newblock \bibinfo{title}{Boundary feedback stabilization of the undamped
  {E}uler-{B}ernoulli beam with both ends free},
\newblock \bibinfo{journal}{SIAM J. Control Optim.} \bibinfo{volume}{43}
  (\bibinfo{year}{2004}) \bibinfo{pages}{341--356 (electronic)}.
\bibitem[{Krall(1989)}]{Krall1989}
\bibinfo{author}{A.~M. Krall},
\newblock \bibinfo{title}{Asymptotic stability of the euler-bernoulli beam with
  boundary control},
\newblock \bibinfo{journal}{Journal of Mathematical Analysis and Applications}
  \bibinfo{volume}{137} (\bibinfo{year}{1989}) \bibinfo{pages}{288 -- 295}.
\bibitem[{Park et~al.(2008)Park, Kang, and Kim}]{Park-Kim2008}
\bibinfo{author}{J.~Y. Park}, \bibinfo{author}{Y.~H. Kang},
  \bibinfo{author}{J.~A. Kim},
\newblock \bibinfo{title}{Existence and exponential stability for a
  {E}uler-{B}ernoulli beam equation with memory and boundary output feedback
  control term},
\newblock \bibinfo{journal}{Acta Appl. Math.} \bibinfo{volume}{104}
  (\bibinfo{year}{2008}) \bibinfo{pages}{287--301}.
\bibitem[{Park and Kim(2005)}]{Park-Kim2005}
\bibinfo{author}{J.~Y. Park}, \bibinfo{author}{J.~A. Kim},
\newblock \bibinfo{title}{Global existence and stability for
  {E}uler-{B}ernoulli beam equation with memory condition at the boundary},
\newblock \bibinfo{journal}{J. Korean Math. Soc.} \bibinfo{volume}{42}
  (\bibinfo{year}{2005}) \bibinfo{pages}{1137--1152}.
\bibitem[{Conrad and Morg{\"u}l(1998)}]{ConradMorgu1998}
\bibinfo{author}{F.~Conrad}, \bibinfo{author}{{\"O}.~Morg{\"u}l},
\newblock \bibinfo{title}{On the stabilization of a flexible beam with a tip
  mass},
\newblock \bibinfo{journal}{SIAM J. Control Optim.} \bibinfo{volume}{36}
  (\bibinfo{year}{1998}) \bibinfo{pages}{1962--1986 (electronic)}.
\bibitem[{Li et~al.(2001)Li, Wang, Liang, Yu, and Zhu}]{Li2001}
\bibinfo{author}{S.~Li}, \bibinfo{author}{Y.~Wang}, \bibinfo{author}{Z.~Liang},
  \bibinfo{author}{J.~Yu}, \bibinfo{author}{G.~Zhu},
\newblock \bibinfo{title}{Stabilization of vibrating beam with a tip mass
  controlled by combined feedback forces,},
\newblock \bibinfo{journal}{Journal of Mathematical Analysis and Applications}
  \bibinfo{volume}{256} (\bibinfo{year}{2001}) \bibinfo{pages}{13 -- 38}.
\bibitem[{Bosello et~al.(2007)Bosello, Lazzari, and
  Nibbi}]{BoselloLazzariNibbi2007}
\bibinfo{author}{C.~A. Bosello}, \bibinfo{author}{B.~Lazzari},
  \bibinfo{author}{R.~Nibbi},
\newblock \bibinfo{title}{A viscous boundary condition with memory in linear
  elasticity},
\newblock \bibinfo{journal}{Internat. J. Engrg. Sci.} \bibinfo{volume}{45}
  (\bibinfo{year}{2007}) \bibinfo{pages}{94--110}.
\bibitem[{Fabrizio et~al.(1994)Fabrizio, Giorgi, and Morro}]{FGM1994}
\bibinfo{author}{M.~Fabrizio}, \bibinfo{author}{C.~Giorgi},
  \bibinfo{author}{A.~Morro},
\newblock \bibinfo{title}{Free energies and dissipation properties for systems
  with memory},
\newblock \bibinfo{journal}{Arch. Rational Mech. Anal.} \bibinfo{volume}{125}
  (\bibinfo{year}{1994}) \bibinfo{pages}{341--373}.
\bibitem[{Deseri et~al.(2006)Deseri, Fabrizio, and Golden}]{DFG2006}
\bibinfo{author}{L.~Deseri}, \bibinfo{author}{M.~Fabrizio},
  \bibinfo{author}{M.~Golden},
\newblock \bibinfo{title}{The concept of minimal state in viscoelasticity: new
  free energies and applications to {PDE}s},
\newblock \bibinfo{journal}{Arch. Ration. Mech. Anal.} \bibinfo{volume}{181}
  (\bibinfo{year}{2006}) \bibinfo{pages}{43--96}.
\bibitem[{Pazy(1983)}]{Pazy}
\bibinfo{author}{A.~Pazy}, \bibinfo{title}{Semigroups of linear operators and
  applications to partial differential equations}, volume~\bibinfo{volume}{44}
  of \textit{\bibinfo{series}{Applied Mathematical Sciences}},
  \bibinfo{publisher}{Springer-Verlag}, \bibinfo{address}{New York},
  \bibinfo{year}{1983}.
\bibitem[{Prato and Sinestrari(1987)}]{DS}
\bibinfo{author}{G.~D. Prato}, \bibinfo{author}{E.~Sinestrari},
\newblock \bibinfo{title}{Differential operator with non dense domain},
\newblock \bibinfo{journal}{Ann. Scuola Norm. Sup. Pisa Cl. Sci}
  \bibinfo{volume}{14} (\bibinfo{year}{1987}) \bibinfo{pages}{285--344}.
\bibitem[{Murakami(1991)}]{Mu}
\bibinfo{author}{S.~Murakami},
\newblock \bibinfo{title}{Exponential asymptotic stability of scalar linear
  volterra equations},
\newblock \bibinfo{journal}{Differential Integral Equations}
  \bibinfo{volume}{4} (\bibinfo{year}{1991}) \bibinfo{pages}{519--525}.

\end{thebibliography}
\end{document}